\documentclass{book}
\usepackage{cimath}

\begin{document}

\chapter*{Les ordinateurs quantiques affrontent le chaos}


\adresse{Laboratoire de Physique Th\'eorique, UMR 5152 du CNRS, IRSAMC,
Universit\'e Paul Sabatier, 31062 Toulouse cedex 04}


\par
\begin{resume}

Le parall\'elisme autoris\'e par la m\'ecanique quantique permet d'effectuer
des calculs d'une mani\`ere radicalement nouvelle.
Un ordinateur quantique fond\'e sur ces principes pourrait r\'esoudre
certains probl\`emes exponentiellement plus vite qu'un ordinateur classique.
Nous discutons comment un ordinateur quantique r\'ealiste
peut simuler une dynamique complexe, en particulier les syst\`emes
chaotiques o\`u les fautes de l'ordinateur classique croissent
exponentiellement vite.

\end{resume}


\begin{multicols}{3}
\section{PROLOGUE}
\par
{\it ``Il se servit un verre de vin blanc, tira les rideaux
 et s'allongea pour r\'efl\'echir.  Les \'equations de la th\'eorie du
chaos ne faisaient aucune r\'ef\'erence au milieu physique dans lequel
se d\'eployaient leurs manifestations; cette ubiquit\'e leur permettait de
trouver des applications en hydrodynamique comme en g\'en\'etique des
populations, en m\'et\'eorologie comme en sociologie des groupes. Leur pouvoir
de mod\'elisation morphologique \'etait bon, mais leurs capacit\'es
pr\'edictives quasi nulles.  \`A l'oppos\'e, les \'equations de la m\'ecanique
quantique permettaient de pr\'evoir le comportement des syst\`emes
microphysiques avec une pr\'ecision
excellente, et m\^eme avec une pr\'ecision totales si l'on renon\c{c}ait
\`a tout
espoir de retour vers une ontologie mat\'erielle.  Il \'etait au moins
pr\'ematur\'e, et peut-\^etre impossible, d'\'etablir une jonction
math\'ematique entre ces deux th\'eories.''} (M. Houellebecq,
{\it Les particules \'el\'ementaires}, Flammarion, 1998).

\section{INTRODUCTION}
\par

Le probl\`eme soulev\'e ci-dessus devient encore plus frappant si
on imagine que la dynamique chaotique est simul\'ee par un
ordinateur quantique.
En effet, depuis une quinzaine d'ann\'ees, la possibilit\'e d'utiliser
la m\'ecanique quantique pour effectuer des calculs
 est de plus en plus discut\'ee dans la communaut\'e scientifique.
Dans les ann\'ees 1980, Richard Feynman a propos\'e d'utiliser des
\'el\'ements quantique pour simuler de mani\`ere efficace le
comportement de syst\`emes quantiques.  En effet, un syst\`eme
quantique \`a $n$ corps \'evolue dans un espace de Hilbert dont la
dimension cro\^it exponentiellement avec $n$. Par exemple,
un syst\`eme de $n$ spins pouvant prendre deux valeurs
\'evolue dans un espace de dimension $2^n$. Ceci oblige un
ordinateur classique \`a effectuer un nombre \'enorme d'op\'erations
pour simuler le comportement de ces syst\`emes m\^eme pour des $n$
mod\'er\'es. Par contre si les op\'erations sont physiquement effectu\'ees
par des \'el\'ements quantiques, il est possible de faire le m\^eme
calcul avec des ressources bien plus petites.  La question se
posait donc de construire une th\'eorie de l'information et une
algorithmique en tenant compte de ces effets quantiques. Ceci a
\'et\'e r\'esolu ces derni\`eres ann\'ees, et on sait maintenant d\'ecrire un
mod\`ele th\'eorique d'information quantique, fond\'ee sur la notion de
``qubits'' (c'est-\`a-dire des syst\`emes quantiques 
\`a deux \'etats $|0\rangle$
et $|1\rangle$) sur lesquels on
agit par des transformations unitaires conservant la probabilit\'e.
 Le qubit est l'\'equivalent
quantique du bit, et d\'ecrit l'unit\'e d'information quantique de la
m\^eme mani\`ere que le bit en information classique.  N\'eanmoins, il
repr\'esente un objet essentiellement diff\'erent, pouvant varier sur
des ensembles continus de valeurs, mais n'en fournissant qu'une en
cas de mesure. De plus, plusieurs qubits peuvent \^etre intriqu\'es,
c'est \`a dire pr\'esenter des corr\'elations entre eux absentes en
information classique.  Un exemple bien connu d'effets de
l'intrication appara\^{\i}t dans le paradoxe Einstein-Podolsky-Rosen.
En effet, pour  deux qubits  dans l'\'etat intriqu\'e
$(|00\rangle + |11\rangle)/\sqrt{2}$, la mesure d'un qubit influera 
l'\'etat de l'autre quelle que soit la distance entre eux.

  Un ordinateur quantique est g\'en\'eralement vu comme
un ensemble de qubits sur lesquels on agit par des transformations
unitaires.  
Ces transformations s'\'ecrivent comme produits de
matrices unitaires \'el\'ementaires appel\'ees portes quantiques et
agissant sur un ou deux qubits
(voir encadr\'e 1).  Le principe de superposition permet alors de cr\'eer
un  \'etat quantique regroupant un nombre exponentiel d'\'etats 
computationnels et d'agir sur eux tous en une seule op\'eration.
Par exemple, pour $n$ qubits initialement dans l'\'etat $|00...0\rangle$,
l'application de $n$ portes d'Hadamard (voir encadr\'e 1)
permet de cr\'eer l'\'etat
$2^{-n/2}\sum_{x=0}^{2^n-1} |x\rangle$ qui regroupe tous les \'etats 
computationnels de $|00...0\rangle$ \`a $|11...1\rangle$.
Des travaux ont permis d'\'etablir que cette ``parall\'elisation''
massive due au nombre exponentiel d'\'etats multi-qubits dans l'espace
de Hilbert pouvait permettre d'acc\'el\'erer de mani\`ere spectaculaire
la vitesse de r\'esolution de certains probl\`emes, d'abord dans la simulation
de la m\'ecanique quantique, mais aussi en d'autres cas comme
la d\'ecomposition en facteurs premiers d'un grand nombre (ceci ayant
des applications fondamentales en cryptographie).  Une tr\`es grande
activit\'e se d\'eveloppe donc dans le monde en ce moment, pour essayer de
construire un tel ordinateur, un probl\`eme
technologiquement tr\`es ardu o\`u plusieurs voies sont
exploit\'ees pour la r\'ealisation physique de
qubits (voir l'article de M.~Brune et J.~M.~Raimond dans ce num\'ero).
\`A pr\'esent des algorithmes simples ont \'et\'e impl\'ement\'es dans
des syst\`emes comportant jusqu'\`a sept qubits  en utilisant les spins
nucl\'eaires de mol\'ecules par r\'esonance magn\'etique nucl\'eaire
(RMN) ou des ions pi\'eg\'es.

L'algorithme le plus c\'el\`ebre est celui d\'evelopp\'e par
Peter Shor qui permet de factoriser un grand nombre en facteurs
premiers exponentiellement plus rapidement que n'importe quel
algorithme classique connu \`a ce jour.  En effet, le
meilleur algorithme classique n\'ecessite un nombre d'op\'erations
qui croit comme $O(\exp[(\log M)^{1/3}(\log \log M)^{2/3}])$
o\`u $M$ est le nombre \`a factoriser, alors que l'algorithme de Shor
requi\`ere $O(n^3)$ op\'erations quantiques \'el\'ementaires o\`u $n$
est le nombre de bits de $M$ ($n \approx \log_2 M$).
Cet algorithme a en fait une grande importance pratique,
car bien que la factorisation soit difficile, l'op\'eration inverse
demande seulement
$O(n^2)$ op\'erations pour reconstruire le nombre
une fois les facteurs connus.  Cette propri\'et\'e d'asym\'etrie
en fait la base
du protocole cryptographique RSA utilis\'e partout dans le monde.
En effet, l'op\'eration facile permet de crypter ais\'ement, 
et l'op\'eration inverse difficile prot\`ege le secret 
du message.
Un ordinateur quantique permettrait donc de briser la plupart
des codes de ce type.  N\'eanmoins, les nombres utilis\'es
par ces codes sont d\'ej\`a tr\`es grands, et un ordinateur
quantique d'au moins $1000$ qubits serait n\'ecessaire
pour les factoriser suivant la m\'ethode de Shor.
\'Etant donn\'e la taille des processeurs quantiques en
construction, il est donc important de trouver des algorithmes
quantiques dont l'utilit\'e se manifesterait pour des nombres de
qubits moins importants, par exemple quelques dizaines.

Un exemple de ce type d'algorithme est repr\'esent\'e par
la simulation de syst\`emes chaotiques, classiques ou quantiques.
En effet, les syst\`emes classiques chaotiques pr\'esentent des
propri\'et\'es d'instabilit\'e exponentielle locale.  Ceci implique
que deux trajectoires initialement
voisines  s'\'eloignent exponentiellement vite l'une de l'autre, ce qui
rend les simulations de tels syst\`emes tr\`es difficiles.  
Pour cette raison, de tels syst\`emes sont 
de bons candidats de probl\`emes pour
lesquels un ordinateur quantique pourrait \^etre utile.
Il est donc
important d'\'etudier la simulation de tels syst\`emes sur des ordinateurs
quantiques.  Un point crucial est \'egalement de comprendre
 comment les erreurs dues aux impr\'ecisions et imperfections exp\'erimentales
affectent la pr\'ecision du calcul quantique, dans le r\'egime
d'extr\^eme instabilit\'e repr\'esent\'e par le chaos classique (cf
prologue).

\section{ALGORITHMES QUANTIQUES}
\par

Un algorithme classique est une suite d'instructions que l'on donne \`a
l'ordinateur et qu'il doit effectuer successivement en manipulant les
bits de sa m\'emoire (qui valent $0$ ou $1$).  Un algorithme quantique
agit de mani\`ere similaire sur des qubits, i.e. des syst\`emes
quantiques \`a deux \'etats $|0\rangle$ et $|1\rangle$.  Si $n$ tels
qubits sont mis ensemble, le syst\`eme total \`a tout instant sera
d\'ecrit par
une fonction d'onde, qui est un vecteur dans un espace de Hilbert de dimension
$2^n$.  L'\'evolution de ce syst\`eme correspond \`a des transformations
unitaires
dans cet espace.  Toute transformation unitaire est licite,
mais pour \'ecrire un
algorithme quantique, on requi\`ere d'\'ecrire les transformations du
syst\`eme en fonction
de transformations \'el\'ementaires appel\'ees aussi portes
(voir encadr\'e 1).  Un algorithme
quantique
sp\'ecifie donc un \'etat initial facile \`a pr\'eparer,
 puis une suite de transformations
\'el\'ementaires (portes) \`a appliquer \`a cet \'etat, et pr\'ecise ensuite
quelle information retirer de la mesure effectu\'ee sur l'\'etat final du
syst\`eme.

L'algorithme de Shor dont il a \'et\'e parl\'e pr\'ec\'edemment suit
cette structure g\'en\'erale.  Il utilise le fait que le probl\`eme de
factorisation peut \^etre ramen\'e au probl\`eme de trouver la p\'eriode
d'une certaine fonction $f$.
Les qubits de l'ordinateur quantique sont regroup\'es en registres,
l'un d'entre eux contenant toutes les valeurs de $x$ de $0$ \`a $N-1$.
Si ce registre contient $n$ qubits, $N=2^n$ et 
$x$ peut ainsi varier de $0$ \`a 
$2^n-1$. $N$ est choisi entre $M^2$ et $2M^2$, 
o\`u $M$ est le nombre \`a factoriser.
Shor a alors montr\'e qu'il \'etait possible de calculer en parall\`ele
toutes les valeurs de $f(x)$ de mani\`ere efficace sur un deuxi\`eme
registre, permettant de passer
de l'\'etat $N^{-1/2}\sum_{x=0}^{N-1} |x\rangle|0\rangle$ \`a
 $N^{-1/2}\sum_{x=0}^{N-1} |x\rangle|f(x)\rangle$. Si on mesure les qubits
du deuxi\`eme registre, on fixe la valeur de $f(x)=u$, et la
nouvelle fonction d'onde est une somme sur toutes les valeurs de
$x$ ayant pour image $u$ par $f$. Mesurer le premier registre \`a ce
stade ne donnerait pas d'information utile. On applique donc une
transform\'ee de Fourier quantique (QFT) (voir encadr\'e 1) \`a cet
\'etat, obtenant une fonction d'onde finale concentr\'ee autour des
multiples de $N/T$, o\`u $T$ est la p\'eriode. La d\'etermination d'un
de ces nombres par la mesure des qubits permet de calculer $T$. Un
\'el\'ement tr\`es important de cet algorithme est l'utilisation de la
QFT, qui peut \^etre effectu\'ee en $O(n^2)$ op\'erations sur un
vecteur de taille $2^{n}$.  La complexit\'e algorithmique de
l'algorithme de Shor, mesur\'ee par le nombre de portes n\'ecessaires
pour factoriser un nombre $M$ comportant $n$ bits ($n \approx \log_2 M$), 
est de l'ordre de
$n^3$. Ce r\'esultat est obtenu gr\^ace \`a l'utilisation de plusieurs
propri\'et\'es sp\'ecifiquement quantiques: la possibilit\'e d'agir sur
tous les \'etats computationnels
 en m\^eme temps (principe de superpositions) puis (par l'action
de la QFT) 
de les faire interf\'erer de mani\`ere constructive (interf\'erences
quantiques). Ceci est impossible sur un ordinateur classique, et
explique pourquoi l'algorithme de Shor est plus efficace que
n'importe quel algorithme classique connu. Il est possible d'impl\'ementer
l'algorithme de Shor sur un ordinateur classique, mais dans ce cas
le nombre d'op\'erations classiques est exponentiellement grand et
l'efficacit\'e quantique dispara\^{\i}t.


\end{multicols}
\par
\begin{encadre}
{PORTES QUANTIQUES}

Un ordinateur quantique est un ensemble de $n$ qubits sur lesquels
on agit par des transformations unitaires bien choisies.  Toute
transformation unitaire de l'espace de Hilbert de dimension $2^n$
peut s'\'ecrire comme combinaisons de transformations locales,
agissant sur seulement quelques qubits, appel\'ees portes
quantiques.  Un ordinateur quantique r\'eel devra permettre
l'impl\'ementation r\'ep\'et\'ee de quelques-unes de ces transformations
\'el\'ementaires, et un algorithme quantique doit fournir la suite de
portes n\'ecessaires pour parvenir \`a l'\'etat final.  Quelques
exemples usuels:

\vskip 0.2cm
-{\bf porte d'Hadamard} s'appliquant \`a un qubit $|0\rangle \rightarrow
(|0\rangle +|1\rangle)/\sqrt{2}$;
$|1\rangle \rightarrow (|0\rangle -|1\rangle)/\sqrt{2}$;

-{\bf porte de phase} s'appliquant \`a un qubit $|0\rangle \rightarrow
|0\rangle$;
$|1\rangle \rightarrow i |1\rangle$;

-{\bf controlled not} ou {\bf CNOT} s'appliquant \`a deux qubits:
$|00\rangle \rightarrow
|00\rangle$;
$|01\rangle \rightarrow |01\rangle$;$|10\rangle \rightarrow |11\rangle$;
$|11\rangle \rightarrow |10\rangle$;
le deuxi\`eme qubit est invers\'e si le premier est dans l'\'etat $|1\rangle$;

-{\bf controlled controlled not} ou {\bf porte de Toffoli} s'appliquant
\`a trois qubits:
le troisi\`eme qubit est invers\'e si les deux premiers sont dans l'\'etat
$|1\rangle$.
\vskip 0.2cm

Certains ensembles de portes sont suffisant \`a eux seuls pour
construire toutes les transformations unitaires: par exemple CNOT combin\'e 
aux transformations \`a un qubit. On parle alors de portes universelles.

\vskip 0.2cm Une partie d'un algorithme quantique s'\'ecrit donc
comme une suite de portes, dont le nombre quantifie la complexit\'e
quantique du processus. Cette notion ne d\'epend pas du choix des
portes \'el\'ementaires, puisqu'elles s'\'ecrivent toutes les unes en
fonction des autres.  Un processus est polyn\^omial si le nombre
d'op\'erations n\'ecessaires est une puissance du nombre de
bits/qubits manipul\'es, exponentiel s'il en est une exponentielle.
On peut v\'erifier que les op\'erations arithm\'etiques apprises \`a
l'\'ecole primaire sont toutes des algorithmes polyn\^omiaux.
L'ordinateur quantique peut de plus additionner ou multiplier
plusieurs nombres en parall\`ele.

\vskip 0.2cm Une transformation unitaire g\'en\'erale est donn\'ee par
une matrice $N{\times} N$ et n\'ecessite $O(N)$ portes \'el\'ementaires.
Cependant, certaines transformations importantes se d\'ecomposent en
un nombre polyn\^omial de portes. Un exemple de telle
transformation, utilis\'ee dans beaucoup d'algorithmes quantiques
est la transform\'ee de Fourier quantique (QFT). Elle utilise $n$
qubits pour transformer un vecteur de taille $2^n$ par:
$\sum_{k=0}^{2^n-1} a_{k} |k \rangle \longrightarrow
\sum_{l=0}^{2^n-1} (\sum_{k=0}^{2^n-1} \exp(2\pi i k l/2^n) a_k)
|l\rangle $. Elle peut s'\'ecrire au moyen des transformations
\'el\'ementaires $H_j$ (porte d'Hadamard appliqu\'ee au qubit $j$)  et
$B_{jk}$ (porte \`a deux qubits appliqu\'ee aux qubits $j$ et $k$ et
caract\'eris\'ee par $|00\rangle \rightarrow |00\rangle$; $|01\rangle
\rightarrow |01\rangle$;$|10\rangle \rightarrow |10\rangle$;
$|11\rangle \rightarrow \exp(i\pi/2^{k-j})|11\rangle$). On peut
v\'erifier que la s\'equence $\Pi_{j=1}^{n} [(\Pi_{k=j+1}^{n}
B_{jk})H_j]$ effectue bien la transformation de Fourier d'un
vecteur de taille $2^n$ en $n(n+1)/2$ op\'erations. \vskip 0.2cm En
pratique, le choix des portes universelles d\'epend de
l'impl\'ementation exp\'erimentale.

\end{encadre}


\begin{multicols}{3}

D'autres algorithmes ont \'et\'e aussi d\'evelopp\'es; en particulier
l'algorithme de Grover permet de faire une recherche dans une base
de donn\'ee non ordonn\'ee de taille $N$, avec un gain quadratique
($O(\sqrt{N})$ op\'erations au lieu de $O(N)$ classiquement).  Des
algorithmes ont \'egalement \'et\'e construits permettant de simuler
efficacement certains syst\`emes quantiques, en accord avec la
suggestion originale de Feynman.

\subsection{Simulation quantique du chaos classique} 

Il peut para\^{\i}tre naturel de pouvoir simuler la m\'ecanique quantique
efficacement sur un ordinateur quantique, bien qu'en fait une
telle t\^ache ne soit pas \'evidente a concr\'etiser sous forme d'un
algorithme pratique.  Il est plus surprenant d'imaginer qu'un
ordinateur quantique puisse simuler efficacement la m\'ecanique
classique. Pourtant, les probl\`emes pr\'esentant du chaos classique
sont tr\`es difficiles \`a simuler sur un ordinateur classique. En
effet, l'instabilit\'e exponentielle conduit \`a une croissance
exponentielle au cours du temps de la moindre impr\'ecision sur la
distribution classique initiale.  Un exemple d'application
chaotique tr\`es connu  est l'application du chat d'Arnold:
$\bar{y}= y+x (\mbox{mod} L); \bar{x} = x +\bar{y}  (\mbox{mod}
1)$, o\`u $\bar{y}, \bar{x}$ d\'esignent les variables apr\`es une
it\'eration. Le mouvement a lieu dans l'espace de phase $(x,y)$ sur
un tore de taille $L$ (entier) dans la direction $y$ et $1$ dans
la direction $x$.  Des r\'esultats math\'ematiques ont permis de
prouver que la dynamique de ce syst\`eme poss\`ede les
caract\'eristiques du chaos tr\`es d\'evelopp\'e, avec par exemple un
exposant de Lyapounov positif $h\approx 1$.  Ceci caract\'erise une
divergence rapide d'orbites voisines dans tout l'espace de phase,
la distance entre elles augmentant avec le temps
comme $\exp (ht)$.
En raison de cette instabilit\'e, les erreurs d'arrondi dues \`a la
pr\'ecision finie d'un ordinateur classique vont se propager tr\`es
vite dans le syst\`eme. Par exemple, pour le Pentium III en double
pr\'ecision l'orbite sera compl\`etement modifi\'ee apr\`es un nombre
d'it\'erations $t= 40$.  M\^eme si la dynamique exacte est r\'eversible,
l'existence de ces fautes coupl\'ees \`a l'instabilit\'e du chaos
d\'etruit la r\'eversibilit\'e de l'\'evolution du syst\`eme dans le temps.

Gr\^ace au parall\'elisme quantique, un ordinateur quantique peut
simuler l'\'evolution d'un nombre exponentiel de trajectoires
classiques en un temps polyn\^omial.  Ceci permet de calculer
de mani\`ere fiable
des quantit\'es globales du syst\`eme en d\'epit de 
l'instabilit\'e exponentielle
qui amplifie tr\`es vite toute impr\'ecision.
Trois registres sont
n\'ecessaires, deux sp\'ecifiant les valeurs de $x$ et $y$ et un
servant d'espace m\'emoire temporaire. En effet, on peut construire
un \'etat initial proportionnel \`a $\sum a_{i,j} |x_i> |y_j> |0>$ o\`u
les $a_{i,j}$ valent $0$ ou $1$, $1\leq i,j\leq 2^{n}$. Il est
possible d'effectuer les deux additions conduisant \`a l'\'etat $\sum
a_{i,j} |2x_i+y_j> |x_i+y_j> |0>$ de mani\`ere parall\`ele en
seulement $O(n)$ op\'erations, en utilisant uniquement $n-1$
qubits dans le troisi\`eme registre pour les retenues.  De cette
fa\c{c}on, $O(2^{2n})$ trajectoires classiques sont simul\'ees en m\^eme
temps par un ordinateur quantique comportant $3n-1$ qubits au total.  
D'autres applications
chaotiques peuvent \^etre simul\'ees de mani\`ere similaire en un nombre
polyn\^omial de portes quantiques. M\^eme une dynamique dissipative
menant \`a un attracteur \'etrange peut \^etre simul\'ee de mani\`ere
efficace. Apr\`es $t$ it\'erations, l'\'etat quantique du syst\`eme
contient les positions des it\'er\'es des $O(2^{2n})$ points
initiaux.  Une mesure de tous les qubits \`a ce stade donnerait
seulement un point de cette distribution, et ferait perdre
l'efficacit\'e quantique.  Pour \'eviter cela, il est possible
d'appliquer une QFT et de mesurer par ce biais des propri\'et\'es
globales d'un nombre exponentiel d'orbites, obtenant ainsi une
information nouvelle qui n'est pas accessible efficacement
classiquement.

Un ordinateur quantique r\'ealiste comportera n\'ecessairement des
impr\'ecisions dues au couplage avec le monde ext\'erieur
(voir l'article de M.~Brune et J.~M.~Raimond dans ce
num\'ero) qui feront que les portes id\'eales seront remplac\'ees par
des portes approxim\'ees qui vont introduire des fluctuations
d'amplitude $\epsilon$ dans les transformations unitaires
associ\'ees. Il est important de comprendre si la dynamique
chaotique va entra\^{\i}ner une augmentation exponentielle de ces
fautes quantiques comme c'\'etait le cas pour les fautes d'arrondi
dans l'ordinateur classique.  La Fig.~1 montre un exemple de la
dynamique donn\'ee par l'application du chat d'Arnold simul\'ee par
l'ordinateur classique et par l'ordinateur quantique.


\par
\centerline{\includegraphics[scale=0.6]{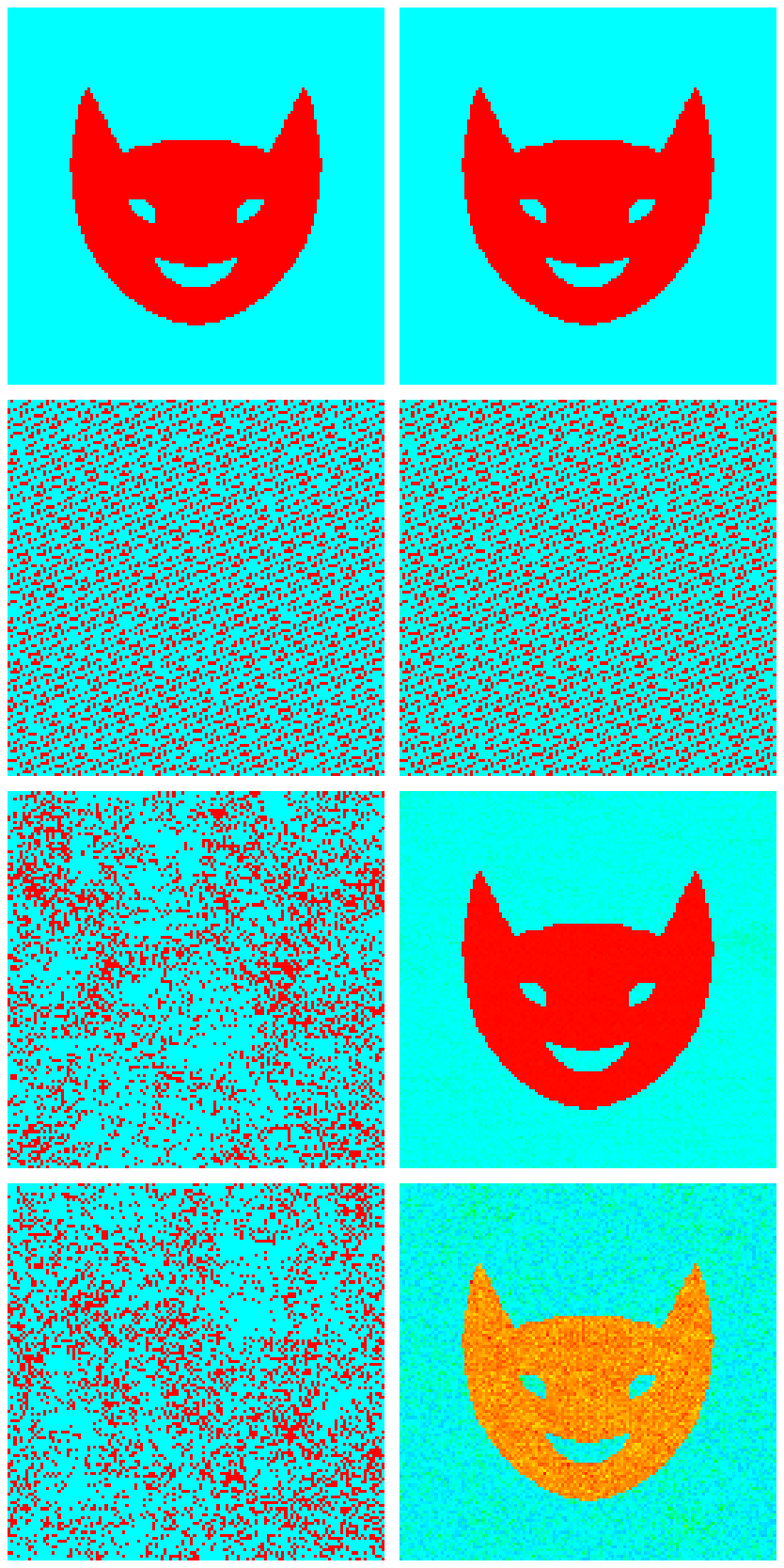}}\nobreak
\legende{Dynamique de l'application du chat
d'Arnold simul\'ee sur un ordinateur classique
(\`a gauche) et quantique (\`a droite), sur un r\'eseau de $2^7 {\times} 2^7$ points
avec une seule cellule ($L=1$ dans l'application du chat d'Arnold).
Premi\`ere ligne:
distribution initiale; seconde ligne: distributions apr\`es
$t=10$ it\'erations;
troisi\`eme ligne:
distributions \`a $t=20$ , avec renversement du temps
effectu\'e \`a $t=10$; derni\`ere ligne:
distributions \`a $t=400$, avec renversement du temps
effectu\'e \`a $t=200$. \`A gauche:
une erreur classique de la taille d'une cellule ($\epsilon=1/128$)
est effectu\'ee au moment du renversement du temps
 uniquement;
\`a droite: toutes les portes quantiques
op\`erent avec des erreurs quantiques d'amplitude
 $\epsilon =0.01$;
la couleur repr\'esente la probabilit\'e
$|a_{ij}|^2$, de bleu (z\'ero) \`a rouge (maximale);  
$n=7$ ce qui fait $20$ qubits au total.}

Dans le cas de l'ordinateur classique, $10$ it\'erations sont suffisantes
pour que la faute minimale sur le dernier bit d\'etruise la r\'eversibilit\'e.
En revanche, dans le cas de l'ordinateur quantique, la r\'eversibilit\'e
est pr\'eserv\'ee avec une bonne pr\'ecision en pr\'esence de fautes quantiques
avec une amplitude comparable.  La pr\'ecision du calcul quantique reste
raisonnable pendant un nombre d'it\'erations $t_f \propto 1/(n\epsilon^2)$.
La raison physique de ce r\'esultat r\'eside dans le fait que chaque
porte imparfaite transf\`ere une probabilit\'e $\epsilon^2$ de l'\'etat
exact vers d'autres \'etats.
Ce r\'esultat souligne la nature tr\`es diff\'erente du comportement
des fautes dans le calcul classique et le calcul quantique.  La th\'eorie
des perturbations quantiques explique la stabilit\'e par rapport
aux erreurs quantiques.  Par contre, les fautes classiques ne rentrent
pas dans ce cadre car m\^eme une tr\`es petite
faute classique est tr\`es grande du point de vue
quantique (renversement d'un qubit), et se propage exponentiellement rapidement
dans le calcul classique comme dans le calcul quantique. Chaque ordinateur a 
donc son type naturel de fautes qui se comportent tr\`es diff\'eremment
dans le cadre de la dynamique chaotique.

Ceci permet de donner un nouvel \'eclairage au vieux probl\`eme de
l'irr\'eversibilit\'e en m\'ecanique statistique, objet de nombreuses
controverses depuis celle entre Loschmidt et Boltzmann au XIXe
si\`ecle. En effet, cette irr\'eversibilit\'e se manifeste malgr\'e le
fait que les \'equations de Newton d\'ecrivant le mouvement
microscopique sont r\'eversibles.  Une l\'egende veut que Loschmidt
ait soulign\'e ce fait \`a Boltzmann en lui demandant ce qui
arriverait \`a sa th\'eorie statistique si on inversait les vitesses
de toutes les particules, pour de cette mani\`ere retourner depuis
un \'etat d'\'equilibre \`a un \'etat initial hors d'\'equilibre. La r\'eponse
de Boltzmann aurait \'et\'e: ``alors essaye de le faire''. Ceci serait
possible sur un ordinateur quantique pour un nombre macroscopique
de particules.  En effet, on peut par exemple utiliser
l'algorithme ci-dessus pour $L\gg 1$.  Dans ce cas, le 
syst\`eme est \'etendu dans la direction $y$ et il est connu que le
chaos
induit une diffusion dans cette direction. 
L'\'evolution
statistique de la distribution initiale de particules est alors donn\'ee
par l'\'equation de Fokker-Planck.  Un exemple de l'\'evolution du
deuxi\`eme moment $<y^2>$ de la distribution est montr\'e dans la
Fig.~2.  M\^eme si la dynamique exacte est r\'eversible, en pr\'esence
de fautes classiques, m\^eme tr\`es petites, la diffusion statistique
reprend rapidement apr\`es le renversement du temps, comme affirm\'e
par Boltzmann.  En revanche, sur l'ordinateur quantique m\^eme
imparfait
 l'\'evolution statistique revient vers
l'\'etat initial hors d'\'equilibre.
Ceci permettrait donc \`a l'ordinateur quantique
d'inverser la fl\`eche du temps pour
des syst\`emes macroscopiques avec
un nombre
de particules pouvant atteindre le nombre d'Avogadro ($6.022 {\times} 10^{23}$)
avec seulement 125 qubits ($L=8$).

\par
\centerline{\includegraphics[scale=0.3]{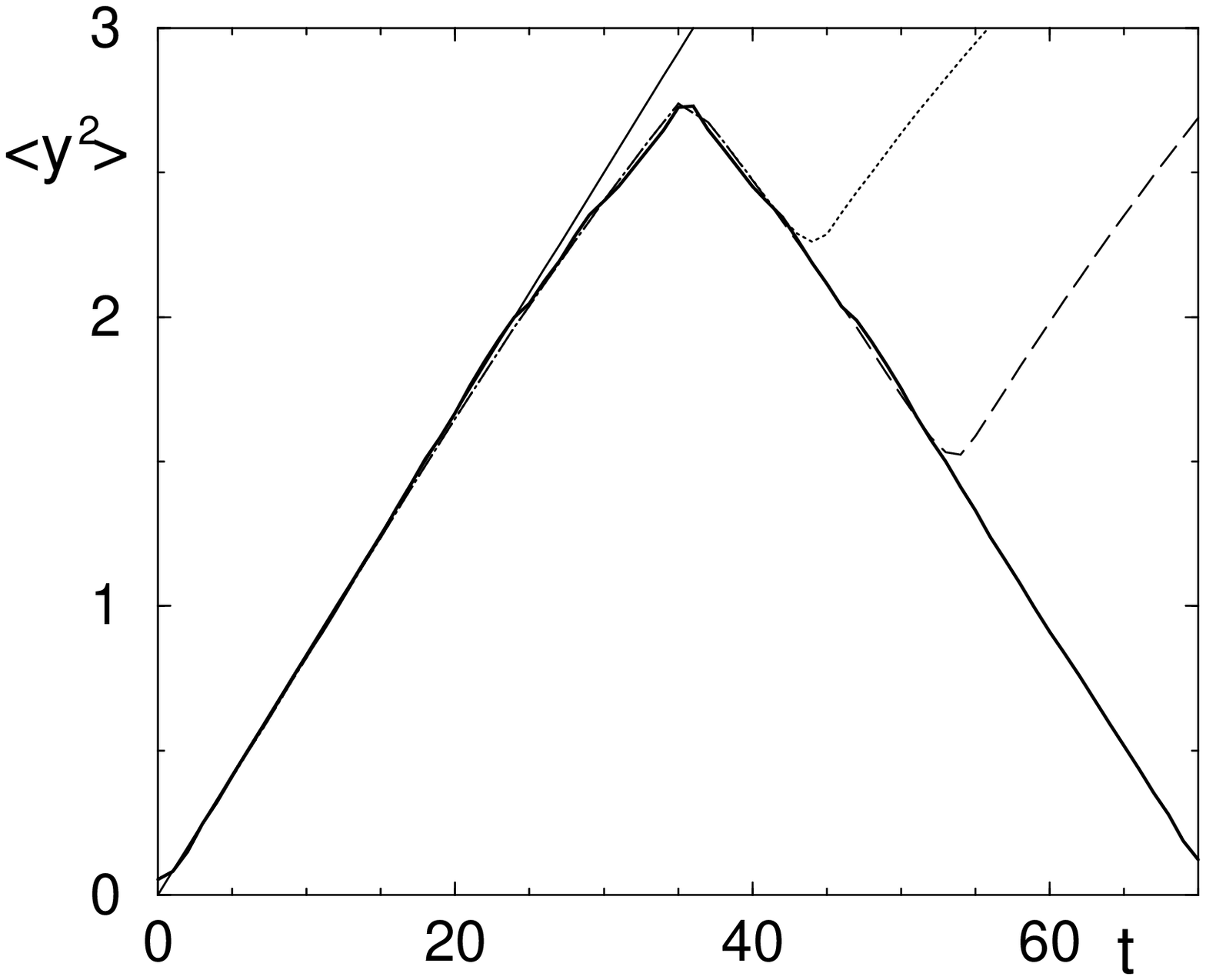}}\nobreak
\legende{Croissance diffusive du second moment $<y^2>$ pour
l'application du chat d'Arnold avec $L=8$, simul\'ee sur des
ordinateurs classique (Pentium III) et quantique (``Quantium I'').
Le fonctionnement de
Quantium I est bien s\^ur lui-m\^eme 
simul\'e au moyen d'un ordinateur classique.
\`A $t=35$ toutes les vitesses sont invers\'ees.  Pour le Pentium III
l'inversion est faite avec une pr\'ecision $\epsilon=10^{-4}$ (ligne
pointill\'ee) et $\epsilon=10^{-8}$ (ligne en tirets); $10^6$
orbites sont simul\'ees. Pour Quantium I, la simulation est
effectu\'ee avec $26$ qubits ($n=7$)(ligne pleine \'epaisse); chaque
porte quantique op\`ere avec du bruit d'amplitude $\epsilon=0.01$.
La ligne droite montre la diffusion macroscopique th\'eorique.}


\subsection{Simulation quantique du chaos quantique} 

Les r\'esultats expos\'es ci-dessus concernent la simulation du
chaos classique.  Il est aussi important d'\'etudier la simulation
de dynamiques quantiques complexes sur des ordinateurs quantiques.
Une classe de syst\`emes particuli\`erement int\'eressante correspond
aux applications quantiques.  Leur dynamique s'\'ecrit de mani\`ere
g\'en\'erale $\bar{\psi}=U \psi$ o\`u $\psi$ est la fonction d'onde
et $\bar{\psi}$ sa nouvelle valeur apr\`es action de l'op\'erateur
d'\'evolution $U=e^{-i\hbar\hat{\ell}^2/2}e^{-KV(\theta)/\hbar}$.
La variable $\theta$
repr\'esente une phase et $\hat{\ell}=-i\partial{}/\partial{\theta}$ correspond
au moment angulaire.
Le param\`etre $\hbar$ repr\'esente la constante
de Planck sans dimension.  Pour de nombreux choix de potentiel $V$,
la dynamique du  syst\`eme classique correspondant devient chaotique
quand $K$ augmente.  Le cas $V(\theta)=\cos\theta$ correspond \`a
l'application standard de Chirikov $\bar{I} = I + K \sin{ \theta };\;
\bar{\theta} = \theta + \bar{I}$, o\`u les barres correspondent
aux nouvelles valeurs de $(I,\theta)$ apr\`es une it\'eration
(l'action classique est $I=\hbar \ell $).
Cette application appara\^{\i}t dans la mod\'elisation
de plusieurs ph\'enom\`enes physiques comme le confinement d'un plasma,
les trajectoires des com\`etes ou le mouvement de particules charg\'ees dans
des acc\'el\'erateurs. Pour $K>0$, ce syst\`eme int\'egrable \`a $K=0$
montre une transition vers le chaos qui suit le th\'eor\`eme de
Kolmogorov-Arnold-Moser.
Pour $K>1$,  la dynamique
classique devient chaotique avec instabilit\'e exponentielle des orbites
et apparition d'une diffusion dans la direction du moment $\ell$
(similaire \`a la diffusion montr\'ee dans la Fig.2).
Le syst\`eme quantique associ\'e, appel\'e rotateur puls\'e,
repr\'esente un mod\`ele tr\`es important
dans l'\'etude du chaos quantique.  Il mod\'elise aussi la physique
des atomes de Rydberg dans un champ micro-onde, permettant des
comparaisons avec les r\'esultats exp\'erimentaux.  De plus, il pr\'esente
un ph\'enom\`ene de localisation d\^u aux interf\'erences
quantiques, par lequel un paquet d'onde quantique
 initialement localis\'e dans l'espace
des moments verra son \'etalement stopp\'e alors qu'une distribution
classique de particules diffuserait dans tout le syst\`eme. Ce ph\'enom\`ene
est similaire \`a la localisation d'Anderson
dans les solides, si bien que son \'etude permet de comprendre \'egalement la
physique des \'electrons dans un syst\`eme d\'esordonn\'e.
Ce mod\`ele a \'et\'e r\'ealis\'e exp\'erimentalement par l'\'equipe
de Mark Raizen \`a Austin (USA) avec des atomes froids.  Le cas
$V(\theta)= (\theta^2-a^2)^2$ correspond \`a un potentiel de double puits
chaotique, et permet d'\'etudier le ph\'enom\`ene d'effet tunnel
quantique assist\'e par le chaos.  En effet, la m\'ecanique quantique
permet \`a un paquet d'onde localis\'e initialement dans un puits
($\theta =-a$ ou $\theta=a$)
de passer d'un puits \`a l'autre p\'eriodiquement.  Ce ph\'enom\`ene
est similaire aux oscillations du chat de Schr\"odinger.  La pr\'esence
de chaos quantique renforce cet effet tunnel de mani\`ere importante,
comme il a \'et\'e vu exp\'erimentalement avec des atomes froids.


\par
\centerline{\includegraphics[scale=0.8]{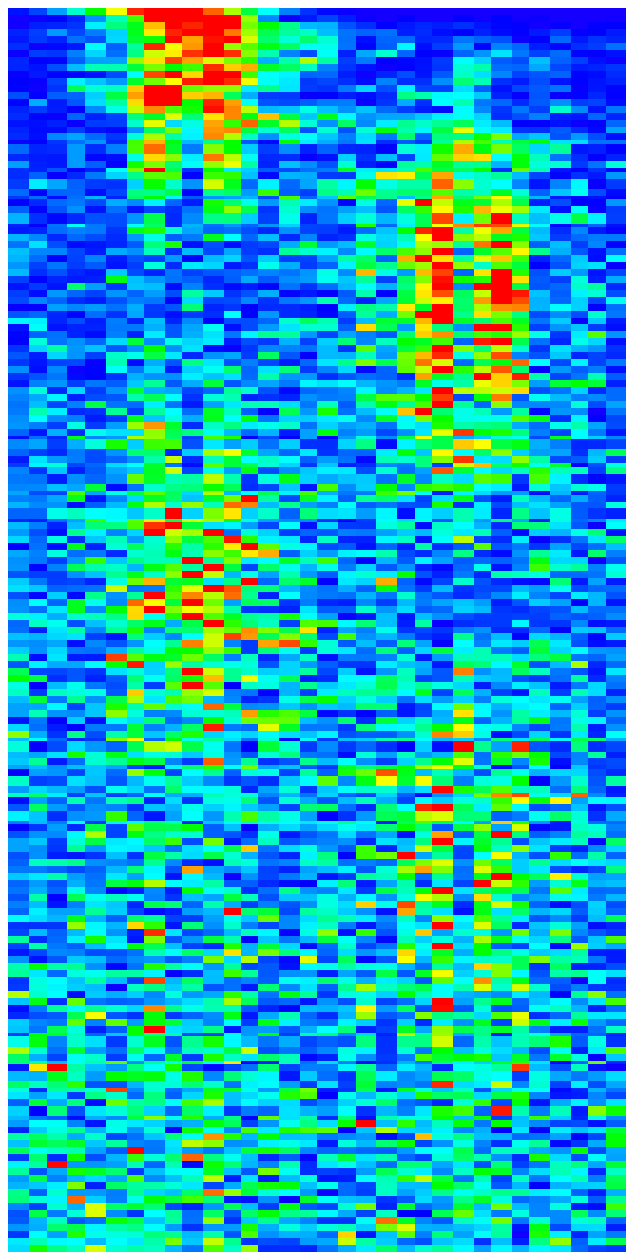}}\nobreak
\legende{\'Evolution du chat de Schr\"odinger (paquet d'onde) anim\'e
par l'ordinateur quantique pour l'application quantique avec
potentiel de double puits.  La probabilit\'e
de pr\'esence en $\theta$ (axe horizontal $-\pi < \theta < \pi$)
est repr\'esent\'ee en fonction du temps $t$ (axe vertical), de $t=0$
(en haut) \`a $t=180$ (en bas).  Les param\`etres sont ici $K=0.04$,
$a=1.6$, $\hbar=4\pi/N$ avec $N=2^{(n-1)}$.  Le calcul quantique
est effectu\'e avec $n=6$ qubits, les portes sont imparfaites avec une
amplitude de bruit $\epsilon=0.02$.  La couleur repr\'esente la densit\'e
de probabilit\'e, de bleu (minimale) \`a rouge (maximale).}


Sur un ordinateur quantique, la dynamique de ces syst\`emes peut
\^etre simul\'ee avec un nombre polyn\^omial de portes pour un vecteur
de dimension exponentiellement large $N=2^{n}$.  Une it\'eration
n\'ecessite $O(n^3)$ op\'erations pour le rotateur puls\'e et
$O(n^4)$ pour le potentiel de double puits, contre $O(N\log N)$
op\'erations sur un ordinateur classique. L'\'el\'ement essentiel de
l'algorithme quantique est l'utilisation de la QFT qui permet de
passer efficacement de la repr\'esentation $\theta$ \`a celle en
$\ell$.

La Fig.~3 montre un exemple d'une oscillation d'un ``chat de
Schr\"odinger'' simul\'ee par l'ordinateur quantique en pr\'esence de
bruit dans les portes. Ce bruit induit une d\'ecroissance des
oscillations suivant une loi en $e^{-\Gamma t}$ avec $\Gamma \sim
\epsilon^2 n^4$ o\`u $\epsilon$
 est l'amplitude du bruit.  La d\'ecroissance \`a taux $\Gamma$
peut \^etre consid\'er\'ee comme une manifestation de la d\'ecoh\'erence,
c'est-\`a-dire une perte de la coh\'erence quantique due au couplage
avec l'ext\'erieur du syst\`eme. La d\'ependance polyn\^omiale de $\Gamma$
dans le nombre de qubits montre qu'il est possible d'effectuer des
simulations de syst\`emes quantiques de tr\`es grande taille sur des
ordinateurs quantiques r\'ealistes.


\section{IMPERFECTIONS DE L'ORDINATEUR ISOL\'E}
\par


Le bruit dans les portes consid\'er\'e ci-dessus correspond \`a une
d\'ecoh\'erence due \`a un couplage avec le monde ext\'erieur. Cependant,
m\^eme en l'absence d'un tel couplage un ordinateur quantique isol\'e
contient des imperfections statiques.  En effet, la distance en
\'energie entre les deux niveaux peut varier d'un qubit \`a l'autre,
dans un intervalle $\delta$. De plus, un couplage r\'esiduel
d'amplitude $J$ entre les qubits est toujours pr\'esent.  En
effet, une interaction entre qubits est n\'ecessaire pour r\'ealiser
les portes quantiques (voir encadr\'e 1). \`A premi\`ere vue il peut
sembler que $J$ est toujours plus grand que la distance $\Delta_n$
entre les niveaux voisins de l'ordinateur quantique isol\'e.  En
effet, la densit\'e de niveaux augmente exponentiellement avec $n$
dans un tel syst\`eme quantique \`a $n$ corps, donnant la loi
$\Delta_n \sim \delta {\times} 2^{-n}$ o\`u $n$ est le nombre de qubits
dans l'ordinateur.  On pourrait penser que pour $J>\Delta_n$ les
niveaux sont m\'elang\'es par l'interaction entre les qubits et les
\'etats propres de l'ordinateur sont fortement modifi\'es par rapport
\`a ceux de l'ordinateur parfait.  Il est clair que limiter les
interactions r\'esiduelles \`a de telles valeurs serait impossible
m\^eme pour quelques dizaines de qubits.  Heureusement, le m\'elange
de niveaux n'appara\^{\i}t en fait que pour des valeurs de $J$ beaucoup
plus grandes, qui d\'epassent le ``seuil de chaos quantique''
$J>\delta/n$. Cette valeur est beaucoup plus grande que
$\Delta_n$ (exponentiellement), et est li\'ee au fait que
l'interaction est \`a deux corps et couple un niveau \`a $n^2$
autres niveaux au plus.  Au dessus de ce seuil, les
caract\'eristiques du chaos quantique apparaissent, un nombre
exponentiel d'\'etats sont m\'elang\'es, les fonctions propres de
l'ordinateur isol\'e deviennent ergodiques, et les niveaux d'\'energie
ob\'eissent \`a une statistique de matrices al\'eatoires.


\par
\centerline{\includegraphics[scale=0.5]{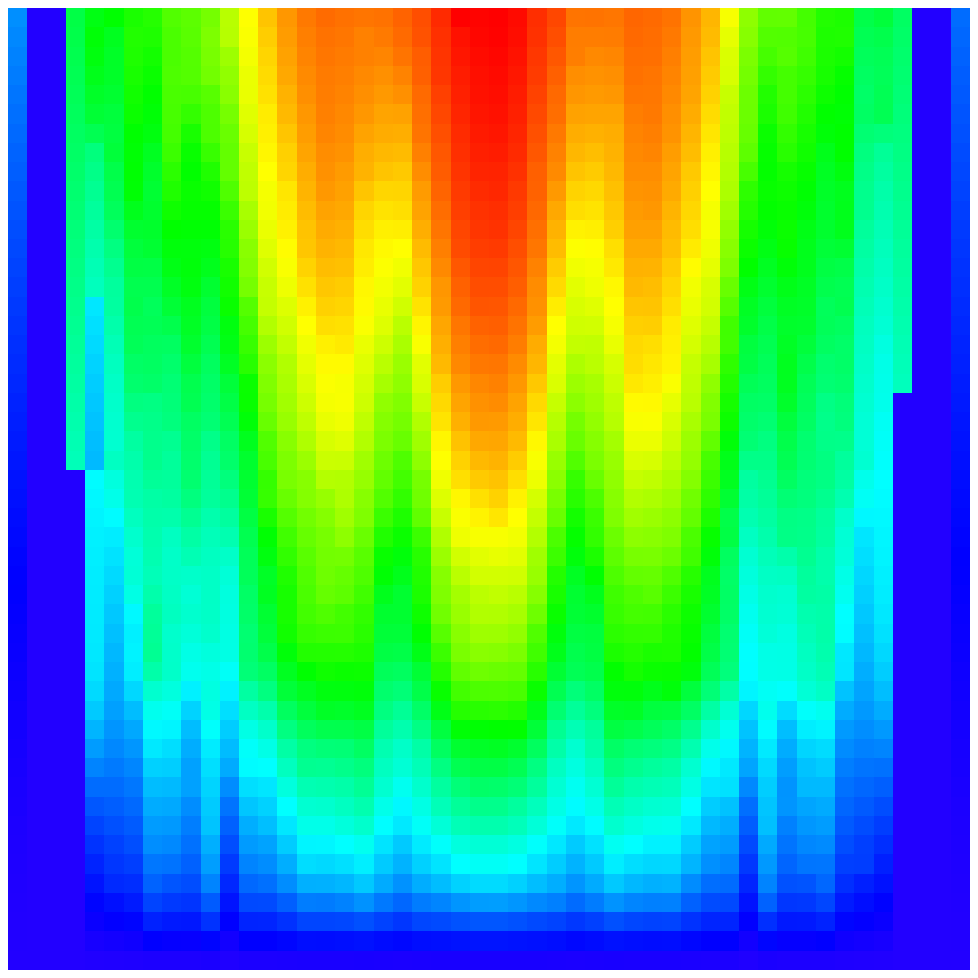}}\nobreak
\legende{Fusion de l'ordinateur quantique induite par le couplage entre
les qubits.  
La couleur refl\`ete l'entropie
des \'etats propres qui caract\'erise le m\'elange des \'etats
parfaits, du bleu (entropie nulle) au rouge (entropie maximale).
Sur l'axe horizontal est repr\'esent\'ee
l'\'energie des \'etats propres de l'ordinateur
compt\'ee
depuis l'\'etat de plus basse \'energie.  L'axe vertical correspond de bas
en haut \`a des
valeurs croissantes du couplage r\'esiduel $J/\delta$, de $0$ \`a $0.5$.
L'ordinateur quantique contient ici 12 qubits.  Le seuil de chaos
correspond \`a $J/\delta \approx 0.3$.}


Un exemple de ``fusion'' de l'ordinateur induit par les couplages
r\'esiduels est montr\'e sur la Fig.~4.  Ce processus commence au
centre de la bande d'\'energie o\`u la densit\'e d'\'etats est maximale et
touche peu \`a peu tout l'ordinateur quand le couplage $J$ augmente.
Il est clair qu'il est pr\'ef\'erable d'op\'erer l'ordinateur en-dessous
du seuil de chaos.  Au-dessus du seuil, le chaos se d\'eveloppe sur
une \'echelle de temps $\tau \sim n \delta/J^2$. Au del\`a de
$\tau$, l'usage de codes correcteurs d'erreur quantiques sera
n\'ecessaire pour continuer le calcul.  De tels codes ont \'et\'e
d\'evelopp\'es, permettant d'effectuer des calculs pour des temps tr\`es
longs en corrigeant au fur et \`a mesure les erreurs, pourvu que
bruit et impr\'ecisions quantiques restent suffisamment petits.

\section{CONCLUSION}
\par

Les r\'esultats pr\'esent\'es montrent que des syst\`emes aussi
complexes que ceux pr\'esentant du chaos peuvent \^etre simul\'es avec
une bonne pr\'ecision sur des ordinateurs quantiques r\'ealistes,
m\^eme avec un nombre mod\'er\'e de qubits, et de mani\`ere beaucoup
plus efficace que sur un ordinateur classique. A pr\'esent, il est possible
de r\'ealiser des ordinateurs quantiques de jusqu'\`a $7$ qubits avec la
technique de RMN.  En principe ceci permettrait de simuler l'\'evolution
pr\'esent\'ee sur la Fig. 3, mais le temps de d\'ecoh\'erence doit encore
\^etre am\'elior\'e pour permettre d'appliquer un plus grand nombre
de portes \'el\'ementaires.


\biblio{POUR EN SAVOIR PLUS}
\rev{Ekert (A.) et Jozsa (R.)}{Rev. Mod. Phys.}{68}{(1996)}{733}
\rev{Steane (A.)}{Rep. Prog. Phys.}{61}{(1998)}{117} \ouv{Nielsen
(M.A.) et Chuang (I. L.)}{Quantum Computation and Quantum
Information}{Cambridge University Press}{(2000)} \rev{Georgeot
(B.) et Shepelyansky (D.L.)} {Phys. Rev. E}{62}{(2000)}{3504}
{Phys. Rev. Lett.}{86}{(2001)}{2890}; {ibid.}{86}{(2001)}{5393}.
\rev{Shepelyansky (D.L.)} {Nobel Symposium on Quantum Chaos Y2K,
Physica Scripta}{T90}{(2001)}{112} \rev{Chepelianskii (A.D.) et
Shepelyansky (D.L.)}{Phys. Rev. A}{66}{(2002)}{054301}


\end{multicols}

\propose{Article propos\'e par: B. Georgeot, tel : 05 61 55 65 63, mail
georgeot@irsamc.ups-tlse.fr, D.L. Shepelyansky,
tel : 05 61 55 60 68, mail dima@irsamc.ups-tlse.fr, page web
http://www.quantware.ups-tlse.fr/
\par Nos recherches ont \'et\'e soutenues en partie par la Communaut\'e
Europ\'eenne dans le cadre du projet EDIQIP de IST-FET et par 
le contrat  NSA/ARDA/ARO No. DAAD19-01-1-0553.}


\end{document}